# DEMONSTRABLY DOING ACCOUNTABILITY IN THE INTERNET OF THINGS

**Lachlan Urquhart, Tom Lodge and Andy Crabtree**

School of Computer Science, University of Nottingham, UK.

Corresponding author: lachlan.urquhart@nottingham.ac.uk

**ABSTRACT**. This paper explores the importance of accountability to data protection, and how it can be built into the Internet of Things (IoT). The need to build accountability into the IoT is motivated by the opaque nature of distributed data flows, inadequate consent mechanisms, and lack of interfaces enabling end-user control over the behaviours of internet-enabled devices. The lack of accountability precludes meaningful engagement by end-users with their personal data and poses a key challenge to creating user trust in the IoT and the reciprocal development of the digital economy. The EU General Data Protection Regulation 2016 (GDPR) seeks to remedy this particular problem by mandating that a rapidly developing technological ecosystem be made accountable. In doing so it foregrounds new responsibilities for data controllers, including data protection by design and default, and new data subject rights such as the right to data portability. While GDPR is 'technologically neutral', it is nevertheless anticipated that realising the vision will turn upon effective technological development. Accordingly, this paper examines the notion of accountability, how it has been translated into systems design recommendations for the IoT, and how the IoT Databox puts key data protection principles into practice.

## INTRODUCTION

The 'connected home' currently sits at the 'peak of inflated expectations' in Gartner's often cited hype cycle,[1] and the Internet of Things (IoT) is a key driver of the hype. A cursory glance at the consumer IoT market reveals swathes of household goods with the prefix 'smart' or 'intelligent' on offer, spanning white good to fixtures and fittings embedded in the fabric of the home.[2] The promise of the IoT is greater convenience, security, safety, efficiency and comfort in a user's everyday life. While the necessity of many IoT products and services may be questionable,[3] anticipated growth in the sector is vast: major IT firms like Cisco, Ericsson, General Electric and Accenture all predict billions of networked devices in the coming years.[4] The IoT essentially trades on data, both actively and passively, with inputs ranging from explicit spoken voice commands to sensed data inputs implicated in such things as movement or temperature monitoring. The IoT also aligns with other trends in computing, particularly big data, cloud computing and machine learning, with personal data collected by IoT devices typically being distributed to the cloud for processing and analytics.

---

[1] Kasey Panetta, "Top Trends in the Gartner Hype Cycle for Emerging Technologies, 2017 - Smarter With Gartner," *Gartner*, 2017, http://www.gartner.com/smarterwithgartner/top-trends-in-the-gartner-hype-cycle-for-emerging-technologies-2017/.
[2] http://iotlist.co
[3] Matthew Reynolds, "These Bizarre Connected Devices Really Shouldn't Exist," *Wired UK*, 2017, http://www.wired.co.uk/article/strangest-internet-of-things-devices.; The Internet of Useless Things Website: http://www.internetofuselessthings.io
[4] Cisco, "The Internet of Everything" (San Jose, 2013), http://www.cisco.com/c/dam/en_us/about/business-insights/docs/ioe-value-at-stake-public-sector-analysis-faq.pdf; Louis Columbus, "Roundup of the Internet of Things Forecasts and Market Estimates, 2016," *Forbes Tech*, 2016, 2, doi:http://www.forbes.com/sites/louiscolumbus/2016/03/13/roundup-of-cloud-computing-forecasts-and-market-estimates-2016/#3a6bfd6e74b0.

Accompanying the diversity of IoT devices and services are concerns centring on privacy and trust. When sensing occurs in the home, for example, patterns of behaviour can be detected and inferences made about inhabitants' lifestyles. Depending who is making these inferences, and who they share the data with, privacy harms can emerge. As Nissenbaum argues, inappropriate flows of information between contexts can cause harm to an individual's sense of privacy.[5] The nascent nature of the industry means there is a lack of harmonised standards for building IoT devices in ways that sufficiently foreground and anticipate data protection concerns.[6] Building trustworthy relationships with consumers in the new IoT infrastructure is critical,[7] and not least because an increasing array of high profile stories about IoT devices leaking data,[8] or being hacked and becoming implicated in widespread distributed denial of service attacks,[9] contribute to a diminishing sense of trust in the emerging infrastructure.

Against this background we elaborate key challenges posed by the IoT from a regulatory perspective and how these practically occasion the need for accountability. These include challenges posed by devices that lack or only provide partial user interfaces and compliant consent mechanisms; the opacity of data flows to end-users and the spectrum of GDPR control rights; machine to machine communications and the legitimacy of access; and cloud storage and international data transfer safeguards. We move on to explore various aspects of the Accountability Principle, first its history in data protection governance and then how it is presented in Article 5(2) of GDPR. This exploration involves questioning the nature of the account to be provided, how it is to be provided, and to whom. We situate Article 5(2) within the wider context of GDPR, turning to various requirements of Article 24 as interpreted in GDPR recitals, and other related Articles, to map how they intersect with accountability. The requirements of GDPR pose distinct challenges to the development of technological systems and we subsequently turn to consider the recommendations of the Article 29 Working Party, and how they envisage GDPR playing out in the IoT, as a preface to presenting the IoT Databox. We conclude by mapping how the IoT Databox addresses the different accountability requirements of GDPR.

**THE PRACTICAL NEED TO BUILD ACCOUNTABILITY INTO THE IoT**
From May 2018, GDPR will be enforced across all European Union member states.[10] It will also affect data controllers outside Europe if they target goods and services to, or otherwise monitor, EU citizens.[11] Seeking to bring data protection laws into the 21st Century, GDPR replaces the pre-Internet Data Protection Directive 1995. The IoT sector is heavily driven by personal data, meaning it is critical that IoT developers negotiate their relationship with the new user rights and controller responsibilities mandated by GDPR. This includes a raft of fresh legal rules governing the processing of personal data, along with extension of the rights provided to data subjects and the responsibilities incumbent on data controllers, all of which are impacted by the underlying technological infrastructure.

---

[5] Helen Fay. Nissenbaum, *Privacy in Context : Technology, Policy, and the Integrity of Social Life* (Stanford Law Books, 2010).

[6] Ian Brown, "GSR Discussion Paper Regulation and the Internet of Things," *International Telecommunications Union* (Geneva, 2015); K Rose, "Internet of Things: An Overview," JOUR, *Geneva: Internet Society*, 2015.

[7] Onora O'Neill TED Talk- https://www.ted.com/talks/onora_o_neill_what_we_don_t_understand_about_trust; UK Digital Catapult Personal Data and Trust Network https://pdtn.org (bring together businesses and academic in network to consider issues surrounding trust in IT) ; Gilad Rosner, *Privacy and the Internet of Things* (O'Reilly Media, 2016), http://www.oreilly.com/iot/free/privacy-and-the-iot.csp.

[8] Richard Chirgwin, 'CloudPets' Woes Worsen: Webpages Can Turn Kids' Stuffed Toys into Creepy Audio Bugs' (*The Register*, 2017)
 <https://www.theregister.co.uk/2017/03/01/cloudpets_woes_worsen_mics_can_be_pwned/> accessed 16 October 2017.Samuel Gibbs, "Hackers Can Hijack Wi-Fi Hello Barbie to Spy on Your Children," *The Guardian*, 2015, https://www.theguardian.com/technology/2015/nov/26/hackers-can-hijack-wi-fi-hello-barbie-to-spy-on-your-children. Cory Doctorow, "Kids' Smart Watches Are a Security/privacy Dumpster-Fire / Boing Boing," *BoingBoing*, 2017, https://boingboing.net/2017/10/21/remote-listening-device.html.

[9] Brian Krebs, "Hacked Cameras, DVRs Powered Today's Massive Internet Outage," *Krebs on Security*, 2016, https://krebsonsecurity.com/2016/10/hacked-cameras-dvrs-powered-todays-massive-internet-outage/. ;

[10] European Commission, "REGULATION (EU) 2016/679 OF THE EUROPEAN PARLIAMENT AND OF THE COUNCIL of 27 April 2016 on the Protection of Natural Persons with Regard to the Processing of Personal Data and on the Free Movement of Such Data, and Repealing Directive 95/46/EC" (Brussels: Official Journal of European Union L119 Vol 59, 2016).

**Lack of or partial user interfaces and consent**
The design of IoT devices is heterogeneous. Unlike mobile phones, where users can develop mental models about how these devices work,[12] 'interfaces' to the IoT vary immensely. Many IoT devices do not have screens and communication with users relies instead on lights or sounds or haptic feedback; text notifications to mobile phones may also be leveraged in the absence of direct device feedback occasioned by the desire to create aesthetically pleasing devices, which may in turn result in opacity about device functionality. This diversity makes it hard for users to understand what personal information is being collected and how it is being used. From a regulatory perspective, this shapes the nature of consent mechanisms. Consent is one legal basis for processing personal data. Consent follows a notice and choice model, meaning it should be informed, unambiguous, freely given and specific to a particular process, and enable a clear indication of the data subject's will.[13] Data subjects need to affirm their choice, and if the type of data being processed is within special categories of personal data (e.g., health, gender, race, or biometric information) explicit consent is needed. Such consent cannot be obtained through pre-ticked boxes, silence or inactivity by the subject.[14] The dominant web-based model takes advantage of the affordances of mobile devices, using screens to display privacy policies, and terms and condition contracts containing large blocks of text. Extensive research shows users do not read this text, as it would take an incredibly long time to do so hence they often agree in any case.[15] Even if they did read it, they cannot renegotiate as it is a form contract and may not understand it due to complex literacy requirements.[16] This situation is not ideal and challenges the notion of legally compliant consent. The heterogeneity of IoT devices could be good or bad for consent processes. On the one hand, consent could be frustrated by devices which, by design, ambiently collect data and have interfaces that lack affordances for communicating clear information. This could be particularly challenging for homes, where children and adults cohabit, as GDPR introduces stricter requirements about delivering clear, concise, comprehensible information to children about data processing.[17] However, on the other hand, the IoT poses an opportunity to redesign how consent is done with users. Taking advantage of new interaction methods may provide for the *ongoing* negotiation of terms of consent.[18]

---

[11] Article 3(2) GDPR

[12] Martina Ziefle and Susanne Bay, "Mental Models of a Cellular Phone Menu . Comparing Older and Younger Novice Users," *Brewster, S., Dunlop, M. (Eds.), Mobile Human Computer Interaction. LNCS 3160. Springer, Berlin, Germany*, 2004, 25–37, doi:10.1007/978-3-540-28637-0_3.

[13] Article 2(h) DPD; Art 4(11) GDPR

[14] Recital 32 GDPR 2016

[15] Aleecia M Mcdonald and Lorrie Faith Cranor, "The Cost of Reading Privacy Policies," *I/S A Journal of Law and Policy for the Information Society*, 2008, http://www.is-journal.org/.

[16] Ewa Luger, Stuart Moran, and Tom Rodden, "Consent for All," in *Proceedings of the SIGCHI Conference on Human Factors in Computing Systems - CHI '13* (New York, New York, USA: ACM Press, 2013), 2687, doi:10.1145/2470654.2481371.

[17] Article 12 GDPR

[18] Ewa Luger and Tom Rodden, "The Value of Consent: Discussions with Designers of Ubiquitous Computing Systems," in *2014 IEEE International Conference on Pervasive Computing and Communication Workshops,*

**Opacity of data flows to end users and control**

IoT devices and the digital ecosystems they feed into are largely opaque in how they handle data. Insofar as end-users may struggle to understand how their devices work, given the lack of effective interfaces, this may in turn lead to lack of legibility in how data is being processed, why, by whom, where it is being stored, for how long etc.[19] This has the knock-on effect of making it hard for users exercise their legal rights and to control use of their information. While no hierarchical framing of rights is encoded in GDPR, a spectrum of various control rights enabling data subjects to escalate action from controllers is nevertheless discernible and underpin accountability in GDPR:

- Article 15 the 'right to access', or the right to discover what data is held by the controller about the data subject.

- Article 16 the 'right to rectification', or the right to correct erroneous data held by the controller.

- Article 21 the 'right to object', or the right to object to the processing of data by the controller.

- Article 18 the 'right to restriction of processing', or the right to require the controller to restrict processing of data.

- Article 20 the 'right to data portability', or the right to have a controller provide data to the data subject in a commonly used, machine readable format to take to another controller.

- Article 17 the 'right to erasure', or the right to have data deleted by controller and for the data subject to thereby be forgotten.

Each of these control rights occasions practical challenges of implementation. If we take data portability, for example, how can data from sensors be moved between IoT service providers in a usable way?[20] Equally challenging and key to control is the need to surface and make visible what information is being processed in the first place.

**Machine to machine communications and access**

The connected home consists of a network of connected devices, many of which may interact with one another. We already see this commercially, with home management system like 'Works with Nest' or Apple's 'HomeKit' linking together manufacture devices and third-party offerings. However, and again due to the paucity of interfaces to the IoT, the lack of human oversight in machine to machine (M2M) communications makes it hard for users to know what is being shared between devices, and if this is contextually appropriate or not. A good example is sensitive personal data collected by a smart mirror detecting someone's skin condition or smart bathroom scales sensing rapid weight loss over time indicating health conditions.[21]

---

Ideally, to respect the agency of users and build their trust, this should not be shared with a health insurance mandated wearable health tracker, unless the user wants it to. Similarly, access by an Amazon Dash inspired replenishment button, perhaps sponsored by a pharmaceutical firm pushing a new skincare range, should have human oversight too. The challenge here is balancing the movement of personal data, utility from devices, business models and ensuring legitimate access to data by different devices and services. By limiting the role of users in the loop, it becomes harder to know if appropriate access is being given (or not) by devices. Linking datasets without adequate access management could also have impacts for data controllers, who need to ensure compliance with DP rules, and users, who may suffer information privacy harms through unexpected data sharing.

**Cloud storage and international data transfer**
The nature of remote, cloud based data storage utilised by most IoT devices is also problematic under GDPR. Services using IoT sensor data often store collected data in servers located outside of the EU. This enables businesses to create large datasets, used in training of machine learning algorithms and finding patterns that can be used either in service delivery, or creation of new services. Managing big data sets raises challenges addressing the velocity, variety, veracity and volume of data.[22] From the perspective of ensuring GDPR compliance, users will struggle to know where their data is, or how they can access and control it when its storage location is likely unknown or geographically distant. Again, oversight over what it is being used for becomes difficult and from a legal perspective, issues of jurisdiction and applicable law in contract clauses can come to the fore. From a data protection stance, adequate protection of data when it leaves the EU is difficult, and measures to guarantee protection, like Privacy Shield (which replaced the former Safe Harbor agreement) or model contract clauses all have their flaws.[23] Furthermore, as mentioned above, Article 3(2) expands the reach of GDPR for controllers outside of the EU monitoring or targeting goods and services towards EU citizens. Cloud providers may not be able to ignore the importance of GDPR in compliance. The alternative of local data storage, keeping information proximate to end users is preferable for ensuring their control over how it is processed, and ensuring more user centric, ethical IoT applications can emerge in the future. The IoT ecosystem, by design, is opaque, and its actions often invisible to end users. In contravention of DP law principles, interactions are being designed that provide little information about how devices function, what data is collected, and what trade-offs consumers are making in order to receive relevant services. This is not sustainable, and risks growth of the sector. It is for these reasons that we argue that accountability needs to be built into the IoT. But what exactly do we mean?

**ACCOUNTABILITY?**
We are of the view that the answer to many of the regulatory issues surfaced by IoT is to build accountability into products and services, by design. Increased dialogue between data controllers and data subjects is needed so that citizens can exercise better control over how their personal data is exploited in the digital economy. Due to GDPR, interest in accountability as a governance mechanism is growing. However, it remains a difficult concept to succinctly pin down. The accountability principle is only substantively mentioned *once* in GDPR, in

---

[22] ICO, "Big Data, Artificial Intelligence, Machine Learning and Data Protection" (Wilmslow, 2017), https://ico.org.uk/media/for-organisations/documents/2013559/big-data-ai-ml-and-data-protection.pdf.
[23] https://www.privacyshield.gov/welcome ; EDPS, Opinion on EU-US Privacy Shield (Brussels, 2016); Article 29 Working Party, Opinion 01/2016 on the EU-US Privacy Shield Draft Adequacy Decision (Brussels, 2016)

Article 5(2).[24] However, its implications quickly spiral when read in the wider context of GDPR, in conjunction with Article 24, various recitals, and other relevant Articles.

Historically, there has been a strong relationship between accountability and data protection compliance. In this context, accountability has traditionally been invoked as a mechanism for implementing data protection principles.[25] As Aldahoff et al. point out,

> *" ... even in instruments where accountability is not called out as a separate data protection principle, many of its substantive provisions were in fact designed to enable accountability".*[26]

The Article 29 Working Party has argued that accountability obliges data controllers to put in place effective policies and mechanisms to ensure compliance with data protection rules,[27] a view endorsed by Aldahoff et al. who underscore the importance of making data processing entities answerable – of 'calling them to account' - for the implementation of appropriate safeguards.[28] The European Data Protection Supervisor (EDPS) argues accountability is not a prescriptive bureaucratic measure merely concerned with validation, but is about proactive leadership to foster a broad culture of accountability.[29] The introduction of GDPR puts measures in place that further develop this culture of accountability.

Adopting a similar framing of the accountability principle created 37 years ago in the OECD Guidelines on the Protection of Privacy and Trans-Border Flows of Personal Data 1980 (paragraph 14), GDPR states,

> *"The controller shall be responsible and be able to demonstrate compliance with, paragraph 1 ('accountability').''* Article 5(2)

This means the controller is responsible for processing personal data in compliance with principles found in GDPR, which are themselves similar to OECD good DP governance principles (paragraphs 7-13). Art 5(1) GDPR includes: (a) lawfulness, fairness and transparency; b) purpose limitation; c) data minimisation; d) accuracy; e) storage limitation; f) integrity and confidentiality. Where OECD and GDPR differ is in the explicit requirement for *demonstration* of compliance with the different principles. Accordingly, there is a two-part responsibility on data controllers: firstly, to put the necessary measures in place to comply with Art 5(1), and secondly, to find ways to demonstrate they have complied. This could be viewed as firstly a 'substantive compliance with principles' requirement, and secondly as a 'procedural demonstration of compliance to relevant stakeholders' requirement. We shall revisit these distinctive aspects of accountability in due course. First we wish to consider what nature an account needs to take and to whom accountability should be demonstrated.[30]

---

[24] Excluding reference in recital 61 and 85 in the context of data breaches
[25] Charles Raab, "The Meaning of 'Accountability' in the Information Privacy Context," in *Managing Privacy through Accountability* (London: Palgrave Macmillan UK, 2012), 17, doi:10.1057/9781137032225_2.
[26] Joseph Alhadeff, Brendan Van Alsenoy, and Jos Dumortier, "The Accountability Principle in Data Protection Regulation: Origin, Development and Future Directions," in *Managing Privacy through Accountability* (London: Palgrave Macmillan UK, 2012), 6, doi:10.1057/9781137032225_4.
[27] EU Article 29 Working Party, "ARTICLE 29 DATA PROTECTION WORKING PARTY Opinion 3 / 2010 on the Principle of Accountability," *A29 Working Party*, 2010, 3, http://ec.europa.eu/justice/data-protection/article-29/documentation/opinion-recommendation/files/2012/wp193_en.pdf.
[28] Alhadeff, Van Alsenoy, and Dumortier, "The Accountability Principle in Data Protection Regulation: Origin, Development and Future Directions," 19.
[29] Ibid., 4.
[30] Bennett, C. (2010) 'International privacy standards: Can accountability ever be adequate?', Privacy Laws & Business International Newsletter, Issue 106, pp. 21–3, p22

**The nature of an account and to whom accountability must be demonstrated**
The current approach in GDPR of not explicitly defining what accountability requires of data controllers is intentional. This again follows OECD 1980 guidelines, which as Alhadeff et al. state,

> " ... do not prescribe to whom the controller should be accountable (the 'accountee'), nor what this relationship should look like."

In their 2010 Opinion on Accountability, the Article 29 Working Party (A29 WP) suggested that putting an explicit accountability principle into GDPR would enable case by case analysis of appropriate measures, and be preferable to predefining requirements due to this approach being more flexible and scalable.[31] Seven years on, if we look to the most recent A29 WP Guidance on Data Protection Impact Assessments, it retains a non-prescriptive stance about measures needed for accountability, beyond publishing DPIAs and the obligation for record keeping. Lack of detailed prescriptive guidance around such a central concept is consistent with original OECD practice, and keeps accountability sufficiently flexible as a notion.[32] Despite the virtues of flexibility, a sticking point for accountability in practice is the form a *demonstrable* account needs to take.

In seeking to answer this, Raab argues that giving an account is akin to 'telling a story' and can be seen to operate at three sequential levels. At its most simple, accountability merely obliges an organisation to report back on its actions. The next level requires mechanisms for that story to be questioned, and for data subjects to offer their own. The third level puts sanctions in place for when an account is poor, either due to inaction or lack of a proper story being offered in the first place.[33] Whilst this provides some abstract navigational aid, it does not pin down the precise dimensions of a good 'account'. A series of European projects including Galway,[34] Paris,[35] and Madrid[36] have been grappling with the nature of accountability. The Paris project document elaborates elements organisations need to put in place to demonstrate accountability. These largely consist of organisational measures, such as establishing policies based on relevant law, setting up internal bodies to enforce these, providing staff training on information privacy, analysing risks on a regular basis, setting up mechanisms to respond to customer complaints, and providing appropriate redress mechanisms. This sits against the wider work of the Galway project, which states accountability in general requires organisational buy in, particularly putting in place internal standards that correlate with external requirements; access to resources to support compliance with policies (training etc.); and internal oversight mechanisms to ensure adherence, coupled with approaches for appropriate sanctions and rule enforcement.[37]

---

[31] Party, "ARTICLE 29 DATA PROTECTION WORKING PARTY Opinion 3 / 2010 on the Principle of Accountability," paras. 44–45; Alhadeff, Van Alsenoy, and Dumortier, "The Accountability Principle in Data Protection Regulation: Origin, Development and Future Directions," 19.
[32] Article 29 Working Party (2017) Guidelines on Data Protection Impact Assessment (DPIA) and determining whether processing is "likely to result in a high risk" for the purposes of Regulation 2016/679, WP248 p12
[33] Raab, "The Meaning of 'Accountability' in the Information Privacy Context," 21.
[34] Link to Galway Project - http://www.informationpolicycentre.com
[35] Link to Paris Project - https://iapp.org/media/pdf/knowledge_center/Accountability_Phase_II.pdf
[36] Link to Madrid Resolution –
http://www.privacyconference2009.org/media/notas_prensa/common/pdfs/061109_estandares_internacionales_en.pdf
[37] M. Abrams, 'Data Protection Accountability: The Essential Elements a Document for Discussion', 2009, 4.

Examining guidance offered by the European Data Protection Supervisor,[38] and UK Information Commissioner Office,[39] we also find a range of new measures in GDPR to assist with accountability requirements. We cluster these in terms of 'technical' or 'organisational' measures:

*Technical Measures*
*Data protection by design and default; including use of anonymization, pseudonymisation and end to end encryption; IT security risk management.*

*Organisational Measures*
*Assigning data protection officers (DPOs); prior consultations; certification schemes; data protection impact assessments (DPIAs);[40] transparent policies; documentation and record keeping on processing for organisations with over 250 staff;[41] internal compliance and audits for effectiveness of approaches; training.*

GDPR thus puts in place a raft of new organisational and technical 'responsibilities' for controllers. Executing these responsibilities is not, as the EDPS puts it, simply a 'box ticking exercise'.[42] The challenge lies in implement these organisational and technical measures as a basis for demonstrating compliance. Thus, it is only through the work of *doing* compliance that accountability comes to life. As Ihde reminds us "Left on a shelf, the Swiss army knife or the cell phone 'does' nothing."[43] The same can be said for the measures mandated by GDPR. It is only when they are built into the everyday practice that complex negotiations between controller and user will emerge, and we can understand what an 'account' may demonstrably look like.

It is equally difficult to succinctly pin down to whom accountability should be demonstrated. The Madrid Resolution attempts to set up international standards on accountability and states that demonstrations should be to supervisory authorities *and* data subjects.[44] However, GDPR is not framed as narrowly. Whilst data subjects and supervisory authorities are clear stakeholders, Article 5(2) is not limited to them, and it is artificial to read Article 5 in isolation of the rest of GDPR, which places many other responsibilities on data controllers. Article 24 specifically focuses on the nature of their wider responsibilities:

*"Taking into account the nature, scope, context and purposes of processing as well as the risks of varying likelihood and severity for the rights and freedoms of natural persons, the controller shall implement appropriate technical and organisational measures to ensure*

---

[38] G. Buttarelli "The accountability principle in the new GDPR" Speech at the European Court of Justice, Luxembourg, 30 September 2016

[39] ICO website - https://ico.org.uk/for-organisations/data-protection-reform/overview-of-the-gdpr/accountability-and-governance/

[40] Article 29 Working Party, "Guidelines on Data Protection Impact Assessment (DPIA) and Determining Whether Processing Is 'likely to Result in a High Risk' for the Purposes of Regulation 2016/679" (Brussels, 2017).

[41] Nature of what should be recorded about processing laid out in Article 30 GDPR. See Art 30(5) on conditions when organisations smaller than 250 persons need to also keep records – e.g. if they are handling special categories of personal data, information relating to criminal convictions etc; See also Recitals 13; 39 and 82 for more detail on reporting.

[42] G. Buttarelli "The accountability principle in the new GDPR" Speech at the European Court of Justice, Luxembourg, 30 September 2016

[43] Don Ihde, "Smart? Amsterdam Urinals and Autonomic Computing," in *Law Human Agency and Autonomic Computing: Philosophy of Law Meets the Philosophy of Computing,* ed. Mirielle Hildebrandt and A Rouvroy (Routeledge, 2011).

[44] Madrid Resolution –
http://www.privacyconference2009.org/media/notas_prensa/common/pdfs/061109_estandares_internacionales_en.pdf

*and to be able to demonstrate that processing is performed in accordance with this Regulation."* Art 24(1) our emphasis

Article 24 surfaces concepts that need to be read in conjunction with Art 5(2), to situate the full extent of data controller responsibilities in GDPR. We focus on the four key issues emphasised above.

**Nature, scope, context and purposes of processing, and risks of varying likelihood and severity.** As these two elements are linked, we consider them side by side. In determining the 'nature, scope, context and purposes of processing', recital 76 GDPR states 'objective risk assessment' is necessary to establish the level of risk attendant to data processing, e.g., if it is risk or high risk. Recital 75 provides examples of particular kinds of risk occasioned by data processing, including when it results in discrimination, financial loss, identity theft or fraud, damage to reputation and reversal of pseudonyms, to name a few. Whilst Article 24 requires assessment of risk, in general, it does not call for a Data Protection Impact Assessments (DPIA) in all cases. However, the focus on risk analysis clearly links to Article 35 which requires a DPIA for processing 'likely to result in high risks'. The nature of 'high risk' is explored in depth in newly released A29 WP guidance, which provides nine examples of high risk processing including processing of data concerning vulnerable data subjects, combining datasets, innovative use of data for new technological/organisational solutions, or preventing data subjects accessing a service.[45] Determining the need for DPIAs, and differentiating the distinctions between risk assessment in Article 24 and Article 35 is a complex exercise. The nine A29 WP examples are quite broad and many IoT applications will likely require a DPIA. This is not necessarily a bad thing, as DPIAs are an important accountability mechanism providing for 'building and demonstrating compliance'.[46] Nonetheless, it is uncertain why Article 24 does not just state DPIAs are always necessary, as this seems to be the practical implication of A29 WP guidance.[47]

**Implementation of appropriate technical and organisational measures**. The language of 'technical and organisational' measures to demonstrate compliance in Article 24 closely aligns with Article 25 requirements on 'data protection by design and default' (DPbD). DPbD obliges data controllers to safeguard the freedoms and rights of individuals at the time of the determination of the means for processing *and* at the time of the processing itself. This may require minimising the processing of personal data, pseudonymising personal data as soon as possible, enabling transparency with regard to the functions and processing of personal data, and allowing the data subject to monitor data processing.[48] In addition, by default, technical and organisational measures should be taken to ensure that:

- Only personal data which are necessary for each specific purpose of processing are processed.

- The amount of personal data collected, the extent of their processing, the period of their storage and their accessibility is controlled.

- Personal data are not made accessible without the individual's intervention to an indefinite number of natural persons.

---

[45] p9-10
[46] p4
[47] p8
[48] Recital 78, GDPR, 2016

In accordance with Article 24, taking into account the nature, scope, context, purposes and risks of processing, DPbD shall reflect the 'state of the art'.[49] This includes putting appropriate *technological* measures in place to demonstrate accountability and achieve compliance. Recital 63 GDPR, for example, states that in regard to data subject rights of access,

> " ... *where possible, the controller should be able to provide remote access to a secure system which would provide the data subject with direct access to his or her personal data.*"

We need to acknowledge, then, that GDPR invokes a turn to the systems design community to engage with data protection challenges, though we acknowledge the nature of design's new, explicit role in data protection regulation remains unsettled.

**Processing is performed in accordance with this regulation.** This requirement brings us full circle back to Article 5(2), that the controller shall demonstrate compliance with accountability, which in turn pulls on other GDPR provisions. The 'lawful, fair and transparent' principle in Article 5(1), for example, requires a turn to Ch II GDPR Articles on lawful processing (Article 6) and consent (Article 7), to name but two. When Article 5(2) is read in context of GDPR as a whole, we also need to examine the nature of data controller responsibility documented in Article 24. Upon doing this, the *breadth* of responsibilities implicated by the accountability principle become apparent. We believe it requires measures for compliance and subsequent demonstration with the entire GDPR. It is hard to isolate provisions of GPDR, as they often connect to and explicitly call on other provisions. This is clear with accountability, which starts as a narrow principle and grows in scope hugely as we dig deeper. Nevertheless, some elements of GDPR align more naturally with the principle. Two examples are transparency (Article 12) and record keeping (Article 30). Thus, accountability turns on the ability to question accounts provided by data controllers around their data handling practices. This requires that record keeping about data processing is in place to demonstrate that compliance with GDPR has been considered and actioned. Similarly, transparency is intrinsically linked to accountability. Transparency mainly focuses on communication by requiring that processing information be provided in clear, concise language which data subjects (and the public at large) can easily comprehend. As framed in GDPR, transparency is less about providing mechanisms to hold controllers to account. Instead it intersects with accountability by dictating the *nature of account giving*.

## ACCOUNTABILITY REQUIREMENTS

Translation of the complex provisions of GDPR into more accessible principles is needed if IoT developers are to build accountability into the IoT. We thus propose 7 actionable accountability requirements, which seek to address key challenges occasioned by the IoT (as outlined above). These requirements highlight manifold 'clusters' of GDPR obligations. This clustering is not exhaustive, but given the broad nature of accountability, we think it provides a useful starting point for considering the nature of an account and substantive elements of GDPR data controllers need to comply with to demonstrate accountability, as outlined below and summarised in Table 1.

### Requirement 1: Limiting Initial Data Collection

GDPR retains the classic data protection principles in Article 5(1) of 'purpose limitation', 'data minimisation' and 'storage limitation'. According to GDPR, personal data should only be collected for 'specified, explicit and legitimate purposes' and not processed in a manner incompatible with those initial purposes.[50] Only what is 'adequate, relevant and necessary' for those initial purposes should be processed.[51] Furthermore, the data should not be kept in a manner which identifies subjects (i.e., not anonymised) longer than necessary for these purposes.[52] Strict oversight over what is being collected, why and how it is managed is necessary.

---

[49] Article 25(1), GDPR, 2016

**Requirement 2: Limitations on International Data Transfer**
GDPR provides strict requirements on when personal data can be sent outside Europe. Article 44 states data should only be transferred to third countries on basis of various conditions. Article 45 states transfers can occur to countries deemed to provide adequate protection by the European Commission, including Uruguay, Israel or New Zealand, to name a few.[53] Other grounds mandate that appropriate safeguards be put in place (Article 46), such as use of standard data protection contract clauses or binding corporate rules that govern data handling in an organisation (Article 47). The Privacy Shield 2016 agreement now covers data transfers to the USA.[54] It requires companies apply the principles of notice and choice, and accountability for onward travel. Minimal oversight is provided by the US Department of Commerce.[55]

**Requirement 3: Responding to the Spectrum of Control Rights**
GDPR provides a spectrum of new control rights around data processing, as described above in Articles 15 to 21. We frame these as rights users can *escalate* as needed from access (Article 15) to rectification (Article 16), objection (Article 21), restriction (Article 18), portability (Article 20) and ultimately erasure and the 'right to be forgotten' (Article 17).

**Requirement 4: Guaranteeing Greater Transparency Rights**
GDPR provides for increased transparency in the relationship between data controller and data subject. Information about processing, particularly concerning data subject rights, is to be provided in:

> *... concise, transparent, intelligible and easily accessible form, using clear and plain language, in particular for any information addressed specifically to a child ... the information shall be provided in writing, or by other means, including, where appropriate, by electronic means.* (Article 12).

Controllers need to furnish the data subject with their identity, the purpose of and legal basis for processing, recipients of their data, and so forth (Article 13). They also need to maintain records of processing under their control, including the actors involved, the nature of processing, type of data collected, security measures taken, and so on (Article 30). The infamous Article 22 also tackles accountability in algorithms and profiling. It provides a right for data subjects not to be subject to decisions based solely on automated processing where the result has significant legal effects (e.g., refusal of credit) or similar (e.g., prejudice from algorithmic profiling).[56] Measures to protect data subjects should be implemented, at minimum, by providing human oversight over such decisions and enabling subjects to voice concerns and contest outcomes.[57] This assumes that the actions and concomitant reasoning of algorithms can be made accountable, a challenge in itself, particularly for machine learning algorithms deployed in conjunction with IoT devices.

---

[50] Art 5(1)(b)
[51] Art 5(1)(c)
[52] Art 5(1)(e) - With exception of longer storage for archiving in the public interest, scientific or historical research or statistical purposes.
[53] REF to current adequacy decisions
[54] It replaces the Safe Harbor Agreement which was deemed inadequate due to the Schrems decision and Snowden revelations about mass surveillance programmes
[55] REF

**Requirement 5: Ensuring Lawfulness of Processing**
Consent is the most discussed grounds for lawful processing of personal data. As discussed in detail above, GDPR provides various requirements for consent mechanisms[58] (see Articles 4, 7, 8 and 9), which are problematic for the IoT. However, consent is not the only basis for lawful processing. Article 6 includes other grounds, namely the legitimate interests of the data controller, the necessity of processing for performance of a contract the subject is party to, or for controller to satisfy a legal obligation they are subject to. Nonetheless, and insofar as the IoT finds its way at scale into consumer goods, consent will remain an important ingredient in ensuring the lawfulness of processing.

**Requirement 6: Protecting Data Storage and Security**
Numerous security and storage requirements exist in GDPR. Accuracy of data is key and appropriate security should be provided, particularly against unauthorised or unlawful processing and against accidental loss, destruction or damage, using appropriate technical or organisational measures (Article 5). This is accompanied by Article 32 requirements to put in place appropriate technical and organisational measures for general security of processing drawing on pseudonymisation and encryption, regular security testing, ensuring resilience of services and timely restoration of data after an incident.[59] GDPR also has strict breach notification provisions around information required and the timeframe for reporting, within 72 hours to authorities (Article 33). For data subjects, what is reported and when is more contingent on severity of breach (Article 34).

**Requirement 7: Articulating and Responding to Processing Responsibilities**
GDPR encourages the adoption of mechanisms for data controllers to articulate their responsibilities. Data Protection Impact Assessments have a key role to play in mapping risks, forecasting their likelihood of occurrence, considering appropriate safeguards, implementing these and making this process of reflection public (Article 35). In highlighting compliance with GDPR, an increased role is envisaged for certification processes, using seals and marks (Articles 42 and 43). Similarly, it is envisaged that new industry codes of conduct will emerge (Articles 40 and 41). In responding to established responsibilities, GDPR guides action by controllers. For organisations of a certain size, an appointed data protection officer will play a key internal oversight and guidance role (Articles 37 and 39). More generally, the turn to technical measures, encapsulated in Article 25 data protection by design and default is *critical* for the IoT.

---

[56] Unless provisions in Art 22(2) apply e.g. its by virtue of a contract, authorised by law or by explicit consent of subject
[57] Extensive literature emerging on existence of a 'right to explanation' and its utility – see Edwards and Veale.
[58] See section above
[59] Art 33 "Taking into account the state of the art, the costs of implementation and the nature, scope, context and purposes of processing as well as the risk of varying likelihood and severity for the rights and freedoms of natural persons". This echoes DPbD provision in Art 25 again.

| Accountability requirement | Source in GDPR |
|---|---|
| 1. Limiting initial data collection | Purpose limitation Art 5(1b); data minimisation Art 5(1c); storage limitation Art 5(1e) |
| 2. Restrictions on international data transfer | Data sent outside Europe on basis of adequacy decision Art 44 and 45; binding corporate rules Art 47; appropriate safeguards Art 46; USA Privacy Shield 2016 |
| 3. Responding to the spectrum of control rights | Right to access Art 15; to rectification Art 16; to object Art 21; to restrict Art 18; to portability Art 20; to erasure Art 17; information supply chain (passing down requests for rectification, erasure, restriction) Art 19 |
| 4. Guaranteeing greater transparency rights | Transparency of information Art 12; rights to provision of information Art 13 and 14; algorithmic profiling Art 22; record keeping Art 30 |
| 5. Ensuring lawfulness of processing | Legality based on specific grounds (Art 5(1a) and Article 6) e.g. performance of contract legitimate interest; consent requirements Art 4 (11), Art 7, Art 8, and Art 9 |
| 6. Protecting data storage and security | Accuracy of data Art 5(1d); integrity and confidentiality Art 5(1f); breach notification to authorities Art 33 and to data subject Art 34; security of processing Art 32 |
| 7. Articulating and responding to processing responsibilities | Articulating responsibilities: Data Protection Impact Assessments Art 35; certifications including seals, marks and certification bodies Art 42 and 43; new codes of conduct Art 40 and 41 Responding to responsibilities: data protection officer Art 37 and 39; data protection by design and default Art 25 |

**Table 1.** Accountability requirements in GDPR.

**IMPLEMENTING ACCOUNTABILITY**

Article 25 introduces data protection by design and default into law, and presupposes not only that technical but *also* technological measures will be put in place to enable demonstrations of compliance with the principle of accountability. However, as GDPR is 'technologically neutral', it offers no guidance to systems designers as to how to build accountability into the technological ecosystem generally or into the IoT specifically. We first consider specific advice from A29 WP for enabling a legally compliant IoT, before moving on to consider the IoT Databox as concrete technological instantiation enabling accountability.

**Technological recommendations for the IoT from A29 WP**
A29 WP provides practical recommendations for IoT developers based on pre-GDPR regulation (DPD 1995), though the principles invoked are incorporated in GDPR. The recommendations seek to foster,

> *"the application of the legal data protection framework in the IoT as well as to the development of a high level of protection with regard to the protection of personal data in the EU."*[60]

A29 WP focuses on three specific IoT developments which directly interface to the user and lend themselves to analysis under data protection laws: wearable computing, such as watches and glasses in which sensors were included to extend their functionalities; Quantified Self, where 'things' are designed to be regularly carried by individuals who want to record information about their own habits and lifestyles; and domotics, IoT devices placed in homes such as connected light bulbs, thermostats, smoke alarms, washing machines, or ovens that can be controlled remotely over the internet. The recommendations are based on perceived challenges to privacy and data protection occasioned by the IoT. These include: lack of user control and information asymmetry, impoverished consent, inferences derived from original data and repurposing of data, intrusive surfacing of behaviour patterns and profiling, limited possibilities to remain anonymous, and security risks. These challenges are considered in relation to data protection law and the requirements that personal data should be collected and processed fairly and lawfully, which requires that data should never be collected and processed without the individual actually being aware of it; the purpose limitation principle, which requires that data is only be collected for specified, explicit and legitimate purposes defined before data processing takes place; that the data collected is kept to the minimum required and is strictly necessary to meet specified purposes, and should not therefore be collected and stored 'just in case' or because 'it might be useful later'; that data is kept for no longer than is necessary for the purposes for which it is collected or is further processed; that sensitive data requires that data controllers obtain the user's explicit consent, unless the data subject has made the data public; that data collection is transparent, which requires data controllers communicate specific information to data subjects concerning the identity of the controller, the purposes of the processing, the recipients of the data, the existence of their rights of access, to withdraw consent and oppose data processing, including information about how to disconnect connected devices to prevent further disclosure of data; that any stakeholder that qualifies as a data controller remains fully responsible for the security of the data processing; and that privacy protections be built-in from the very outset through application of the 'privacy by design' principle. Pre-empting GDPR, consideration of data protection law also includes the right to portability, which is invoked in a bid to foster service switching, to unlock competition impediments, and foster innovation.

---

[60] A29 WP223 (2014) Opinion 8/2014 on Recent Developments on the Internet of Things, http://ec.europa.eu/justice/data-protection/article-29/documentation/opinion-recommendation/files/2014/wp223_en.pdf

The upshot of these considerations is a raft of practical recommendations for IoT developers, including general recommendations 'common to all stakeholders' and more specific recommendations for OS and device manufacturers. General recommendations include conducting Privacy Impact Assessments (PIAs) before any new applications are launched in the IoT; applying the principles of Privacy by Design and Privacy by Default; empowering users to exercise their rights and thus be 'in control' of their data at any time according to the principle of self-determination of data; informing users of their rights in terms that are understandable to them and are not confined to general privacy policies on controllers' websites; designing devices and applications to inform people (users and non-users) as to data collection and processing (e.g., via a device's physical interface or by broadcasting a signal on a wireless channel); deleting raw data as soon as controllers have extracted the data required for their data processing and doing so as a matter of principle at the nearest point of data collection (e.g., on the same device after processing).

Specific recommendations for OS and device manufacturers include implementing Security by Design from the outset; notifying users to update devices when security vulnerabilities are discovered or when a device will no longer be updated; informing users about the type of data collected by sensors and how it will be processed; providing granular choice over data collection when granting access to applications, including choice over categories of data and the time and frequency at which data are captured; offering 'do not collect' options to schedule or quickly disable sensors; providing settings that enable different individuals using the same device to be distinguished and to prevent them learning about each other's activities (interpersonal accountability); limiting the amount of data leaving devices by transforming raw data into aggregated data directly on the device; enabling user rights, including the right of access and data portability by providing user-friendly interfaces that allow users to easily export their data (aggregated or raw) in a structured and commonly-used format; users should also be able to exercise their right to withdraw consent and have this communicated to all stakeholders; furthermore, to *enforce* transparency and user control, device manufacturers should also provide tools to locally read, edit and modify the data before they are transferred to any data controller; the implementation of *local controlling and processing entities* should allow users to have a clear picture of data collected by their devices and facilitate local storage and processing without having to transmit the data to the device manufacturer.

**Enabling local controlling and processing: the IoT Databox**

The IoT Databox puts the local control recommendation into effect to enable the principle of self-determination.[61] The IoT Databox is an 'edge' solution, which is to say it sits at the edge of the network in the user's home, and thus moves computing to the data rather than data to centralised computing as currently prevails under the auspices of 'the cloud'. It is a both a storage and gateway device, enabling applications to access online sources of data (e.g., social media) and to collect data from physical devices (e.g., smart appliances) situated in the domestic environment. The IoT Databox thus collates and mediates third party access to personal data generated within the home *and* data distributed online across multiple silos, which may also include APIs to proprietary IoT devices.

The IoT Databox is predicated on the 'Dataware model',[62] which sought to develop a business to consumer (B2C) service-oriented architecture providing a new wave of personal digital services and applications to individuals. The model posits a *user* (by or about whom data is created), *data sources* (e.g., connected devices or online accounts, which generate or contain data about the user), a *personal container* (which collates the data produced by data sources and can be accessed via an Application Programming Interface or API), a *catalogue* (which allows the user to manage access to the personal container), and *data processors* (external machines exploited by data controllers who wish to make use of the user's data in some way).

The Dataware model is a logical entity formed as a distributed computing system. Data processing involves *requests* being sent to the catalogue, which are approved or rejected by the user. If approved, the catalogue issues a processing *token* to the data processor for permitted requests. The processor presents the token to the personal container, which accepts the token, runs the processing request on the relevant data sources, and then *returns* processed results to the data controller. The Dataware model represents a distinctive approach to personal data processing, which limits the amount of data leaving connected devices and aggregates data at the nearest point of data collection (i.e., at the edge of the network) to enable data minimisation. Furthermore, insofar as data distribution is limited to the results of data processing, then the raw data is not transferred and remains under the user's control.

The IoT Databox embeds the Dataware model in a networked mini-computer, which can be situated in the home and placed under the direct control of its users. It exploits a Security by Design approach at the outset, storing data in manifold containers rather than one container to minimise the potential attack surface and security problems associated with general purpose operating systems.[63] Data processing is done through apps which, like data stores, run within isolated containers on-the-box and interact with data stores to perform a specified (purposeful) task. Apps may query data stores, write to a connected device's store to perform actuation, and/or write to a communications data store that sends query results to external machines. Data stores record all queries, actuations, or external transactions performed on them in an audit log, and access to data stores is rigidly enforced by an 'arbiter' component, which mints and manages the use of access tokens.

---

[61] Perera, C., Wakenshaw, S., Baarslag, T., Haddadi, H., Bandara, A., Mortier, R., Crabtree, A., Ng, I., McAuley, D. and Crowcroft, J. (2017) "Valorising the IoT Databox: creating value for everyone", *Transactions on Emerging Technologies*, vol. 28 (1), http://doi.org/10.1002/ett.3125

[62] McAuley, D., Mortier, R. and Goulding, J. (2011) "The dataware manifesto", *Proceedings of the 3rd International Conference on Communication Systems and Networks*, pp. 1-6, Bangalore, IEEE, 2011.

[63] Mudhavapeddy, A. and Scott, D. (2014) "Unikernels: the rise of the virtual library operating system", *Communications of the ACM*, vol. 57 (1), pp. 61-69.

Architecturally the IoT Databox (Fig.1) consists of three key components: the *Databox* (a networked mini-computer), an *app store* (of which there may be many), and data controller/third party *processors* (again, of which there may be many).

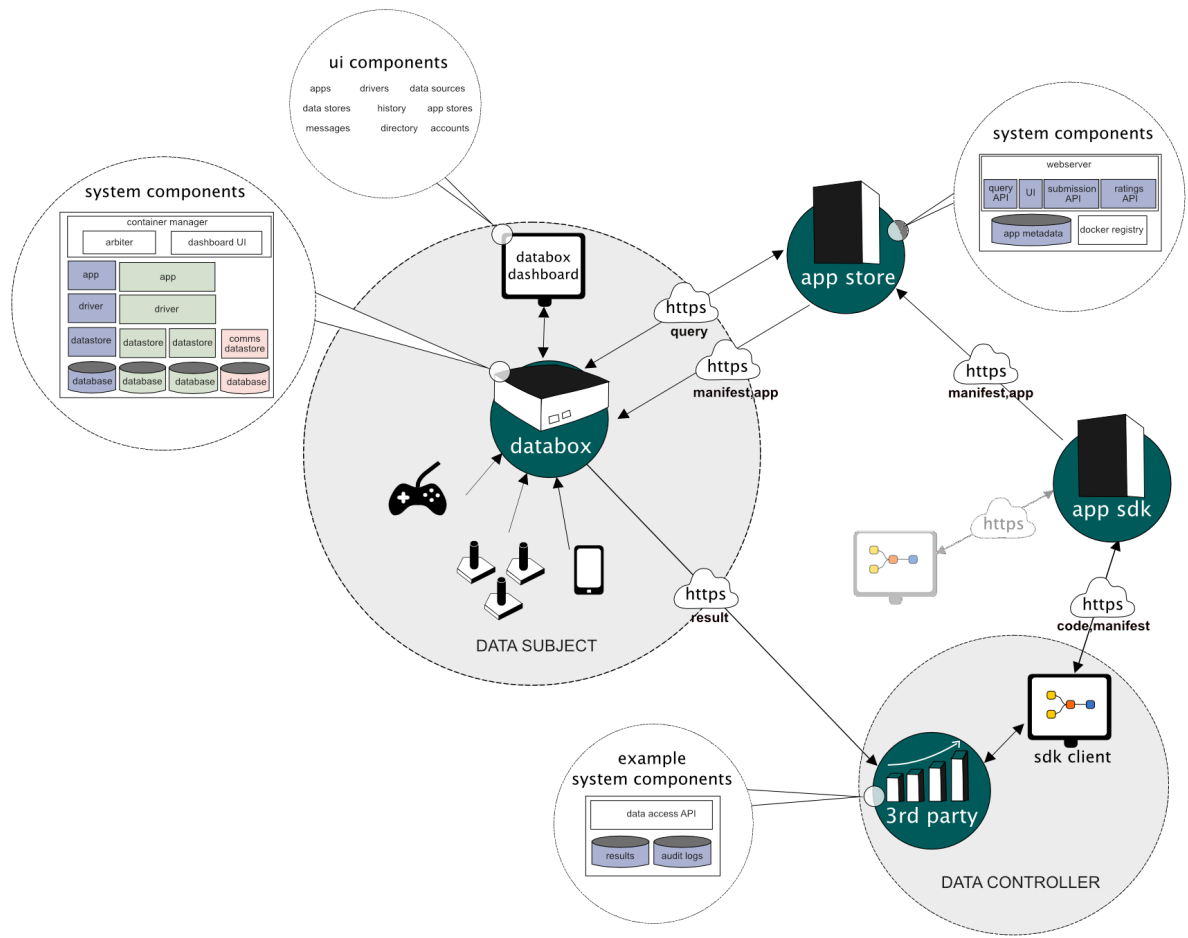

**Figure 1.** The IoT Databox Model.

Users interact with the IoT Databox via the *Dashboard* (Fig.2), which provides users with a range of management functions including:

- Creating *User Accounts* on the Databox and activating sharing permissions (e.g., that consent from all users of shared resources is required for delete actions).
- Adding *Data Sources* to the box; including assigning ownership to data sources, annotating data sources (e.g., smart plug X is 'the kettle'), and sharing data sources with other Databox users.
- Configuring *Drivers* to enable data sources to write to data stores.
- Managing *Data Stores*; including sharing stores with other Databox users, and redacting, clearing, or deleting stores.
- Accessing *App Stores*; apps are recommended by the box based on available data sources but individuals can also search for, download, and rate apps.

- Sharing *Apps*, with other users within the home and between distributed Databoxes in other homes; the Dashboard also allows apps to be updated and deleted.
- Receiving *Notifications*; including the results of data processing prior to distribution, sharing requests, app and device driver updates, resource contention, etc.
- *Auditing* data processing operations; including viewing all accesses to data stores and data transactions, and enabling data processing to quickly put on hold or terminated.

The accounts model and sharing options built into the Dashboard provide the resources needed to enable users to manage one another's individual and collective access to personal data sources within the home.

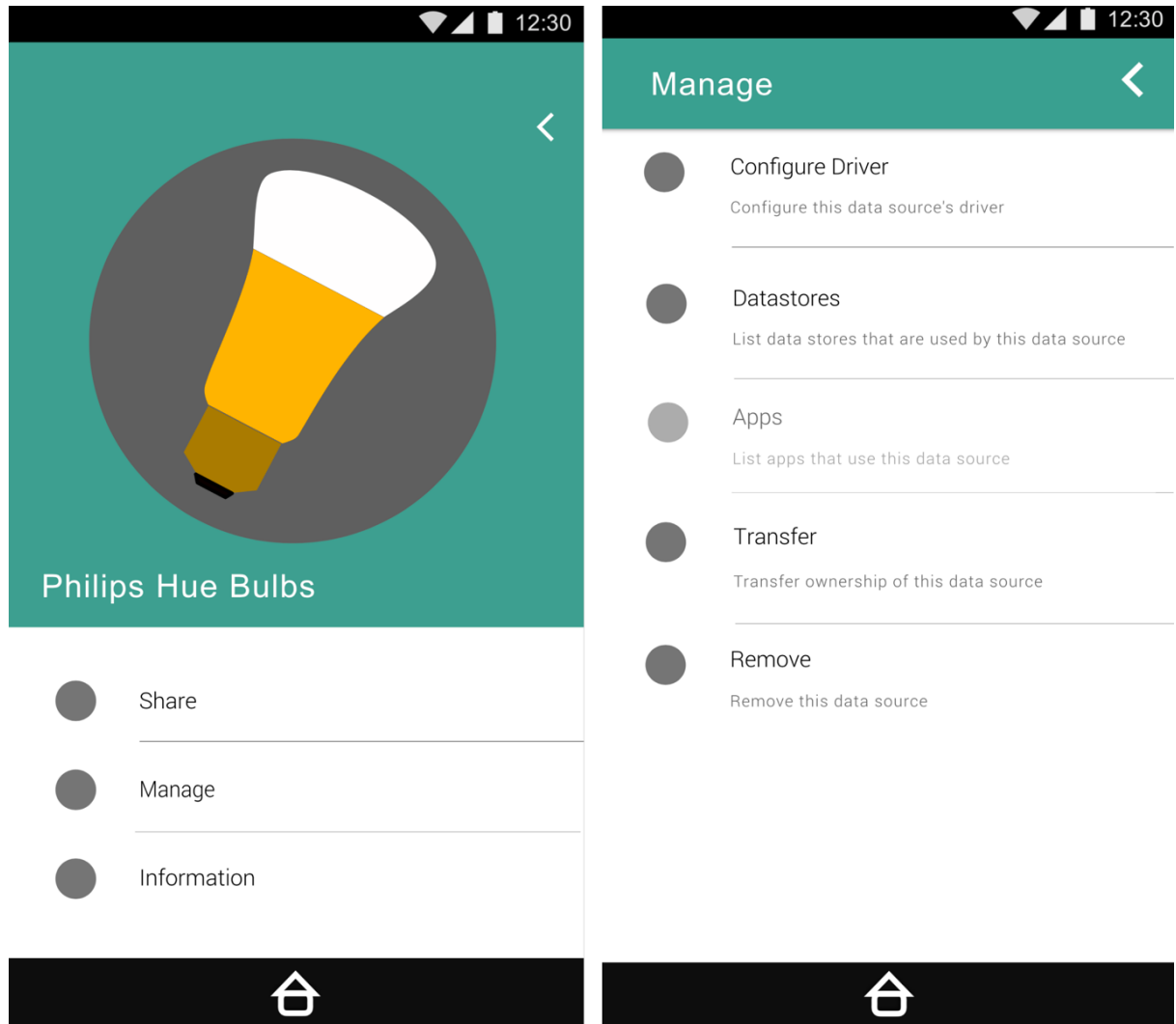

**Figure 2**. The IoT Databox Dashboard.

The *app store* is secure cloud-based service, which manages a repository of apps and app metadata including user ratings, risk ratings, and IoT Databox accreditation for apps that do not take data off-the-box. As noted above, there may be many instances of an app store, including specialisations (e.g., connected home or quantified self app stores). While app developers are free to create their own containerised apps, the app store provides a dedicated Software Development Kit (SDK) supporting the app building and publication process. The app SDK utilises IBM's Node Red flow-based programming paradigm,[64] which connects 'nodes' together to create applications. There are three principle node types: *data sources*, *processes* (functions that operate on data; they typically have a single input connection and one or more output connections) and *outputs* (which typically perform an action, such as actuation, visualisation, or data export). Figure 3 depicts an app that takes the output from a microphone, performs some processing on the data and updates a visualisation, turns on one or more bulbs, and exports the processed data to the cloud. It is composed of a single *data source* (yellow node), three *processes* (blue nodes) and five *outputs* (orange nodes).

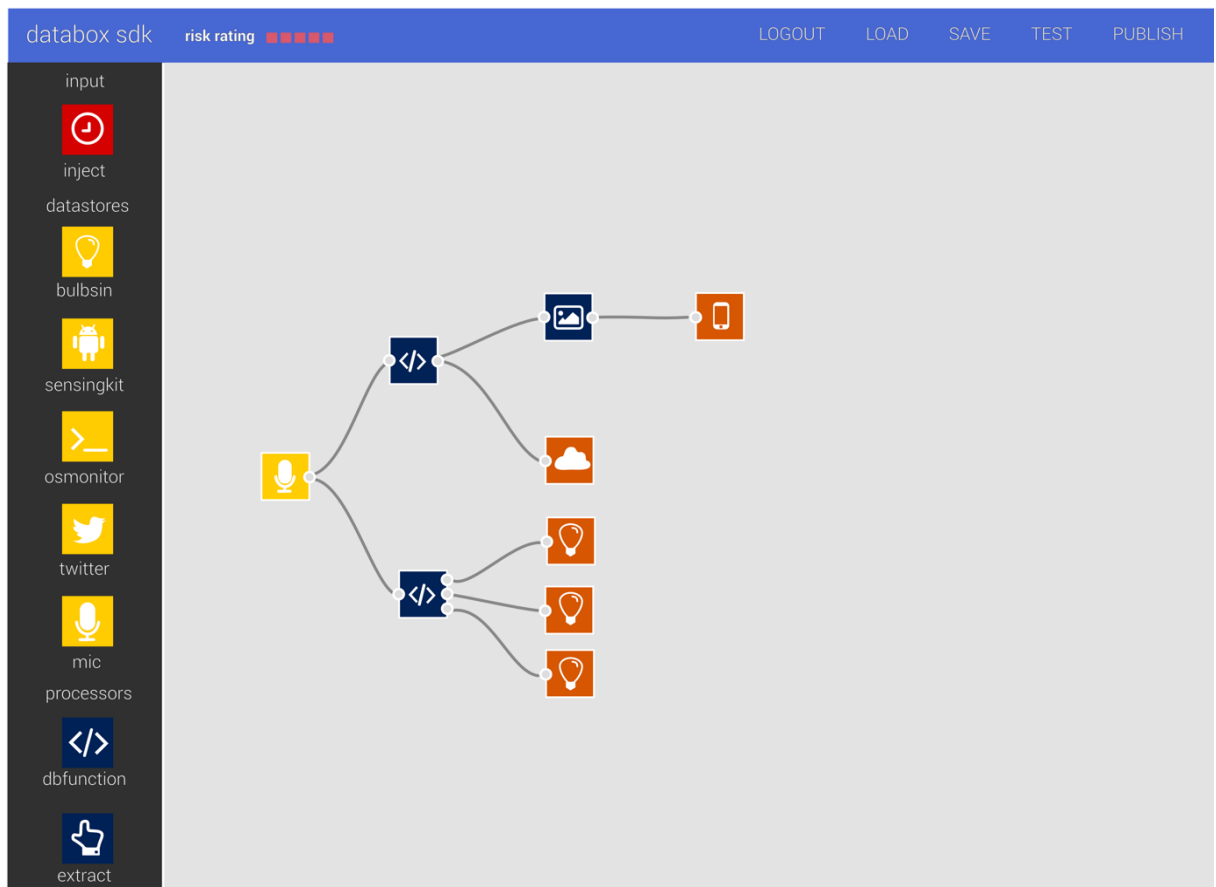

**Figure 3.** The App SDK.

The app SDK smooths and simplifies the build-test-deploy workflow; it presents a high level abstraction (e.g., an app developer can build an app without needing to be familiar with the interoperation between specific sources, stores and drivers); it provides 'scaffolding' to help build an app (e.g., developers can quickly inspect the structure and type of data entering and exiting a node); it provides a full testing environment, where flows are deployed (as containers) and connected to test data; it handles the app publication process; and, upon submission, it containerises an app and uploads it to the app store. The SDK also takes care of source code management as all stages of the app development cycle are recorded in a developer's GitHub account.

Importantly the SDK also seeks to sensitise app developers to the potential risks that accompany personal data processing. We differentiate between three types of risk: *legal risks* associated with GPDPR, particularly those implicated in taking data off-the-box including data export within the EU, outside the EU, transfer to other recipients, the provision of adequacy decisions or safeguards, and access; *technological risks*, including apps that use devices that have not been validated by the SDK, use unverified code, or physically actuate essential infrastructure or potential dangerous devices in the home; and *social risks*, including apps that access sensitive information or produce results that may be deemed sensitive (as articulated, for example, by the notion of 'special categories' of personal data in Article 9 GDPR).We take the view that app developers should be clear about the nature and level of risk of posed by an app and provide precise information about the risks they potentially expose users to.

---

[64] https://nodered.org

We appreciate that identifying risk is challenging, given that it can be introduced by any individual component of a system (both hardware, such as sensors/actuators, and software, such as drivers and apps) as well as arbitrary combinations of the two in particular operating contexts. Though by no means infallible, the SDK generates a risk rating for apps, based on the aggregate risk of the nodes from which it is composed. Each node in the development environment has a pre-defined spectrum of risk attached to it. The final risk rating assigned to the node will sit within this spectrum, and will be determined by how nodes are configured (e.g., the hardware they work with, the proposed data rate, the particular actuation to be performed, etc.). The SDK provides developers with a view on potential risk *in the course of* app construction (Fig.4) in a bid to reduce risk in the IoT ecosystem. The risk rating of apps is also made available to users on the app store (Fig.5) to further motivate and drive the development of low risk and even no risk apps that do not export data, provide users with granular choice over data sampling and reporting frequency, provide online access if apps take data of the box, clearly flag that they actuate essential infrastructure in the home (e.g., central heating or windows and doors), and exploit accredited hardware and trustworthy software. Low-risk apps are visibly 'checked' in the app store to display their Databox accredited status.

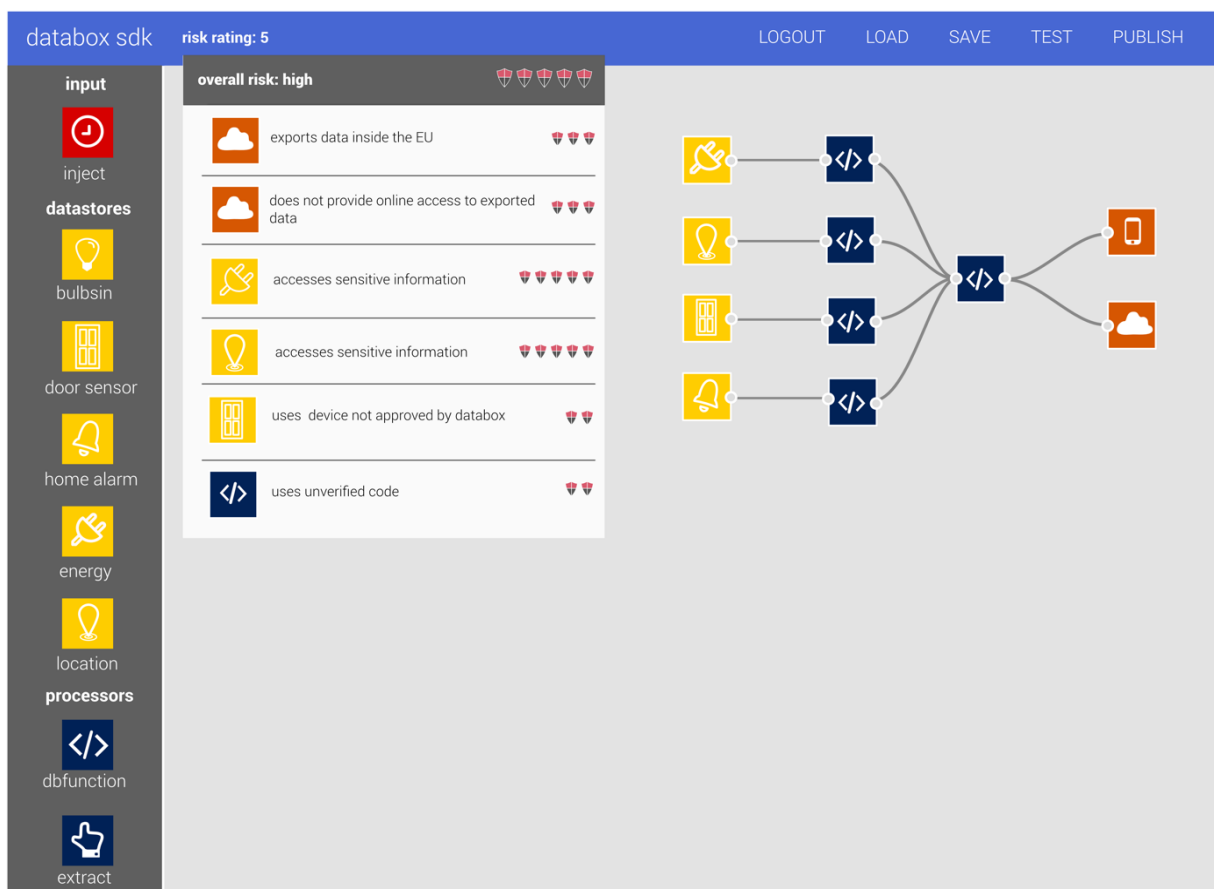

**Figure 4.** SDK risk rating apps during development.

The risk rating assigned by the SDK is reflected in the app store once uploaded. For apps built outside the SDK, the app store reviews and rates them based on features and information provided, e.g., the absence of an API providing users with access to their data would result in a high-risk rating if data was taken off-the-box by an app. Apps may also be posted on the app store with an 'unverified' status, in which case their risk rating will also be high. However, an app cannot be posted on the app store or installed on the IoT Databox without a *manifest* being in place, and data (i.e., the results of processing) cannot be transferred to a controller's processors without a manifest being completed by the individual or data subject.

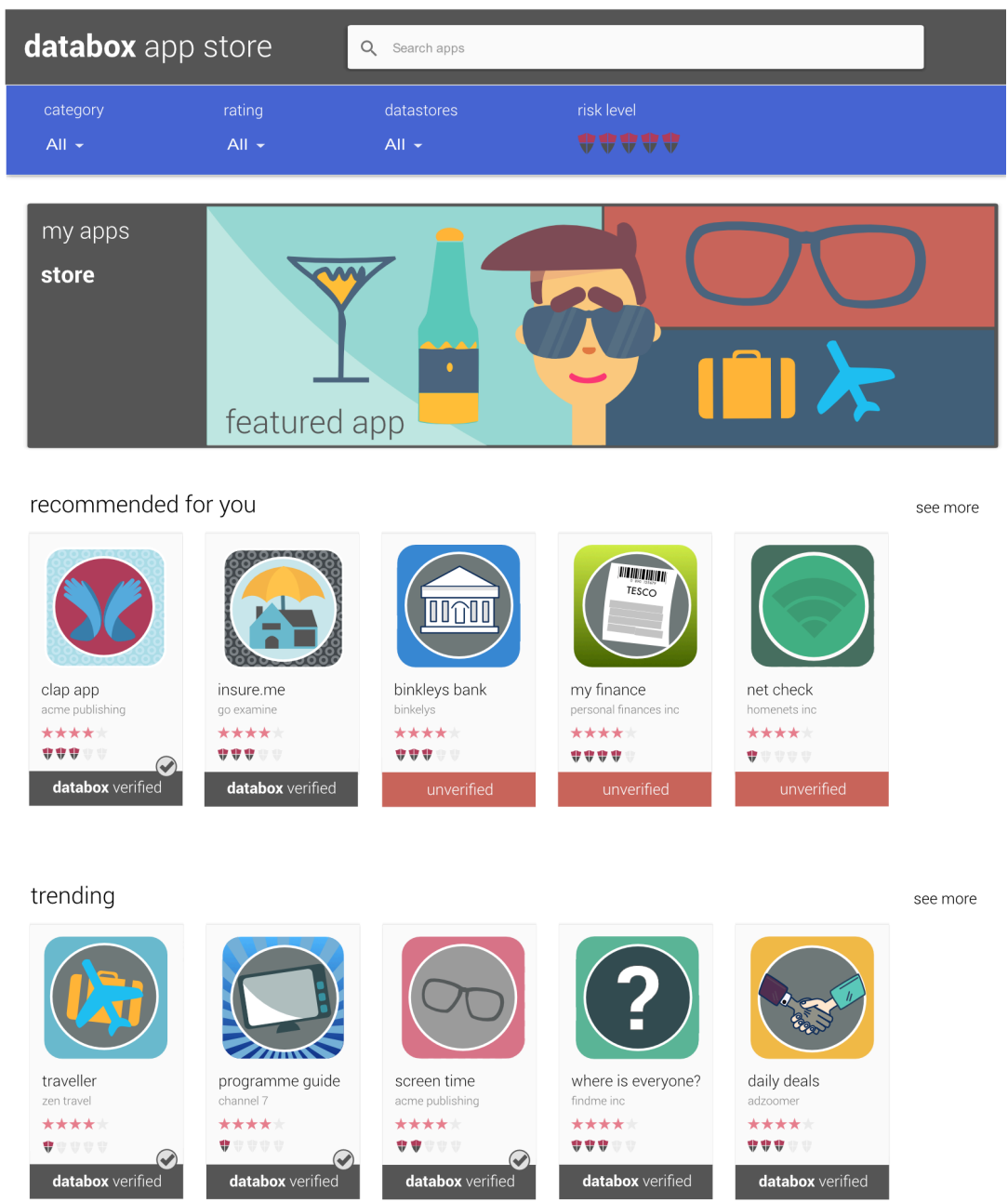

**Figure 5.** At-a-glance risk (sheilds) and user ratings (stars).

Manifests are at their most basic 'multi-layered notices'.[65] They provide a) a *short* description of the specific purpose of data processing, b) a *condensed* description furnishing the information required under Article 13 GDPR, and c) *full* legal terms and conditions. The IoT Databox also adds *app information* to the short description, including an app's risk profile, user ratings and its verified status, and enables *control* over data collection at device level. The IoT Databox thus transforms multi-layered notices into dynamic, user-configurable *consent mechanisms* informing potential app users as to the specifics of data collection and processing, enabling granular choice over categories of data and the time and frequency at which data are captured, and articulates the individual's rights on-the-box rather than on a remote website (Fig.6).

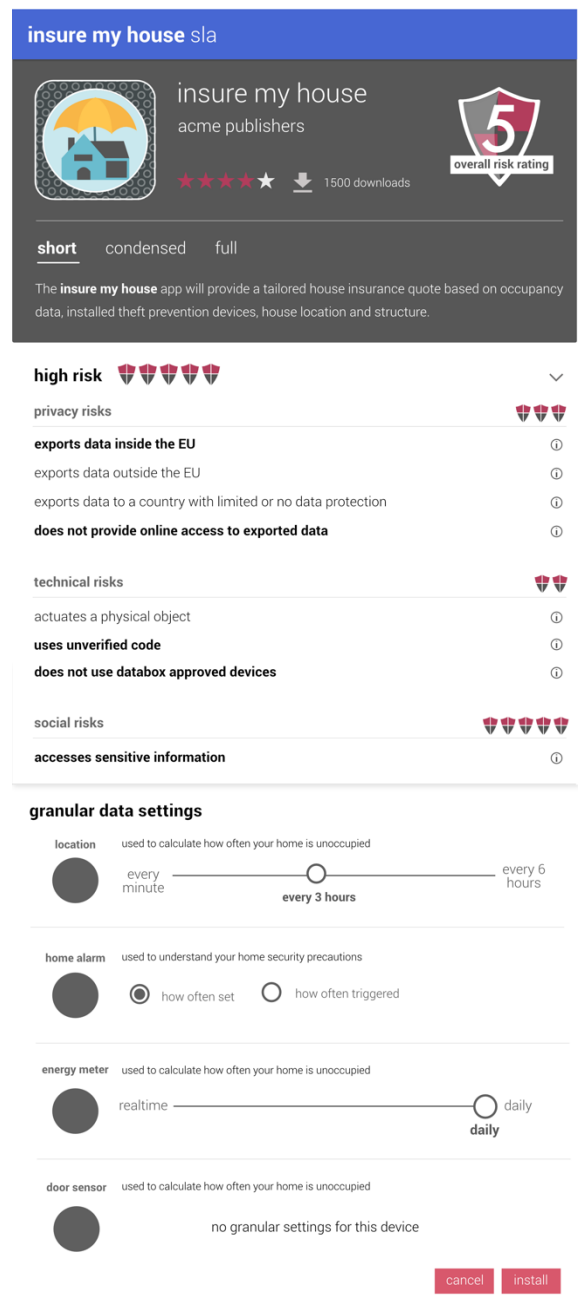

**Figure 6.** Manifest enabling granular choice.

---

[65] A29 WP202 (2013) "Opinion 02/2013 on apps and smart devices", *Article 29 Data Protection Working Party*, http://ec.europa.eu/justice/data-protection/article-29/documentation/opinion-recommendation/files/2013/wp202_en.pdf

The short layer of a manifest provides a placeholder for a 'plain language' purpose description. It provides a description of the risks that attach to using an app independently of the app developer. It describes the data sources an app might use and also allows users to select just which data sources it will use insofar as choice exists. The manifest also allows users to determine the frequencies at which data will be gathered, *and* reported where data is taken off the box. Once configured by the individual, and an app is installed, the manifest assumes the status of a Service Level Agreement (SLA), which the IoT Databox transforms into a set of machine readable policies the arbiter draws upon to *enforce* a data processor's access to the particular data sources agreed upon by the individual and to regulate subsequent data processing operations.

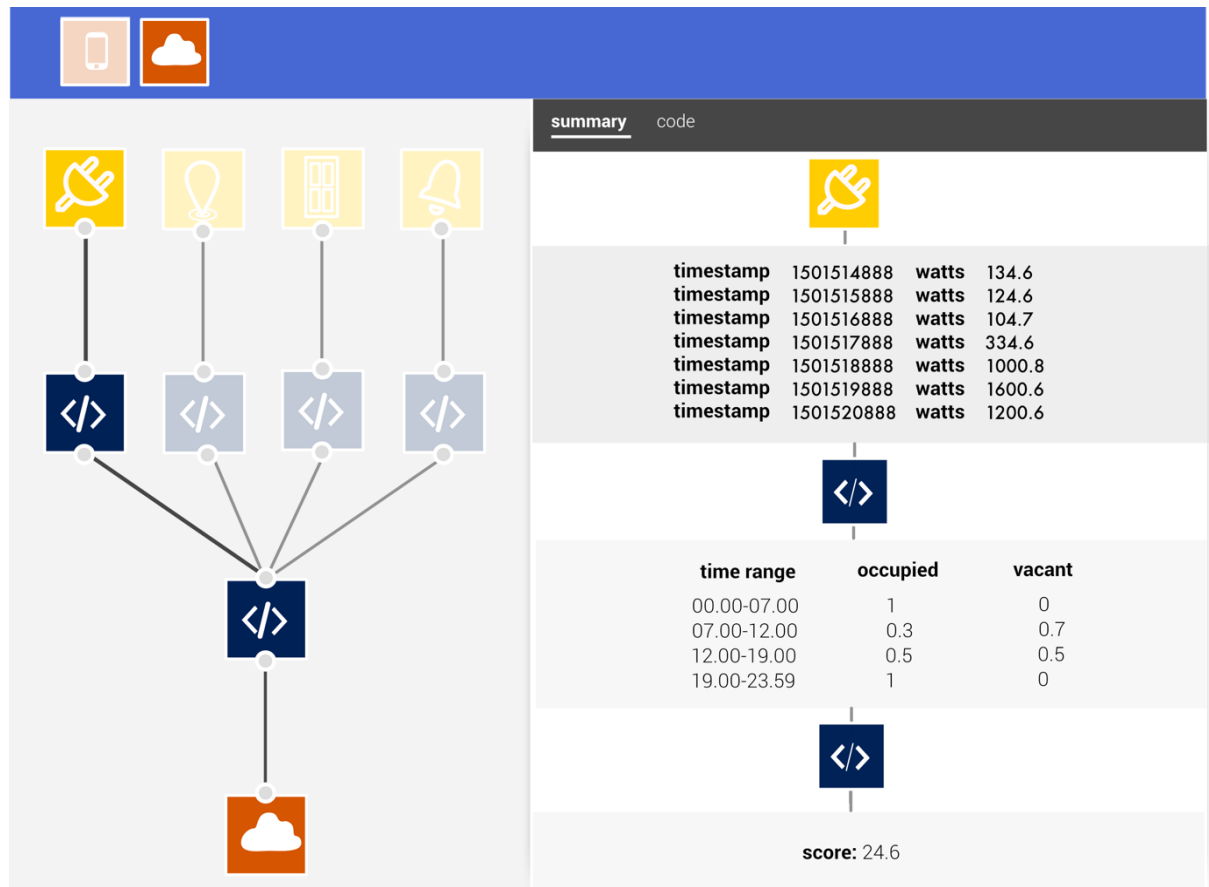

**Figure 7.** Building runtime accountability into apps.

As more and more connected devices find their way into the home, and an increasing array of apps consume personal data and operate on users' behalf, we expect the ability to inspect *what* has happened and *why* will become a necessary feature of app usage. For example, I know my health insurance app provides quotations based on my activity, grocery shopping, location, and financial data, but just how has it arrived at the quotes that it does? Alternatively, one might wonder why the radiators in the living room were set to the maximum at 3AM yesterday, or why a large order of toilet roll has appeared on the doorstep? Whatever the particular case, GDPR makes it clear that the logic, significance and consequences of automated processing should be made accountable to individuals. This may in part be provided in the information contained in consent mechanisms as a preface to app use but, as the above examples indicate, there is also a need to build *runtime* accountability into the IoT.

To enable runtime accountability, and in addition to dashboard notifications, apps created in our SDK are bundled with an *inspection* interface that surfaces *how* an app 'operates', i.e., how data flows through an app and how some action or decision is arrived at, in order to support real-time interrogation by users. By way of example, figure 7 illustrates how data is processed as it moves along the flow path. The path summarises how energy data is used as part of a calculation of a final score sent back to a third party to generate a home insurance quote. The timestamp and watts listing displays the raw data from the energy data source. When it is subsequently processed by the first function node (in blue) it is transformed into an occupancy matrix for times of the day, with the values for house 'occupied' and 'vacant' represent probabilities in the range from 0 to 1. Finally this data, alongside data from other data sources implicated in the other flow paths (location, alarm and door sensor data), is provided to the final function to produce an overall score. As with our attempt to convey potential risk, this is a nascent first step towards enabling runtime accountability. Nonetheless we think it an important area of research and topic of future work, particularly with respect to how an app's operations are conveyed to users, given the emphasis placed on automated processing by GDPR.

**A unique approach?**

We are of course not the first to espouse the virtues of Privacy Enhancing Technologies (PETs). In particular, a raft of Personal Data Stores (PDS) or Personal Data Management Systems (PDMS) have emerged over recent years. Many provide users, like Mydex,[66] with encrypted data stores distributed across the cloud against which a wide a variety of third party applications can be run. Despite the phenomenal growth in such solutions - the WEF reports that more than one a week was launched between January 2013 and January 2014 alone [67] – widespread public uptake has been problematic. Ironically, a recent report suggests that this is due to "perceptions of privacy and security risks" individuals attach to storing their personal data *in the cloud*.[68]

Alternatives are provided by solutions such as openPDS and HAT. OpenPDS is hosted on either a smartphone or an internet-connected hard drive situated in the home.[69] OpenPDS provides users with a centralised location for storing personal data and exploits the 'SafeAnswers' approach to compute third-party queries inside a software sandbox within the user's PDS returning, like the IoT Databox, only the results of processing not the raw data.[70] HAT provides users with a personal container that also stores data client-side.[71] Purpose-built 'data plugs' fish personal data from APIs and deposit it into a user's personal HAT container. HAT-enabled applications can then access a user's data through 'data debits', which permit access to raw data in return for specific services. The primary purpose of HAT is to create a *marketplace* that redresses the current asymmetry in data harvesting and builds users into the personal data value chain.

The MyData initiative takes a different approach again. It does not provide a PDS solution, but instead seeks to enable *consent management*.[72] MyData thus provides a digital service that focuses on managing and visualising data use authorisations, rather than storing data itself. It seeks to encourage service providers to build MyData APIs, which enable their services to be connected with MyData accounts. MyData APIs enable interaction between distributed data sources and data users, and the MyData account provides users with a single hub for granting services the authority to access and use their personal data. While the MyData account lets individuals activate or deactivate the sharing of specific data flows and lists currently active authorisations, it does not put further measures in place to limit access and minimise data distribution.

---

[66] https://mydex.org
[67] World Economic Forum (2014) *Rethinking Personal Data: A New Lens for Strengthening Trust*, http://www3.weforum.org/docs/WEF_RethinkingPersonalData_ANewLens_Report_2014.pdf
[68] Larsen, R., Brochot, G., Lewis, D., Eisma, F. and Brunini, J. (2015) *Personal Data Stores*, https://ec.europa.eu/digital-agenda/en/news/study-personal-data-stores-conducted-cambridge-university-judge-business-school
[69] de Montjoye, Y., Wang, S. and Pentland, A. (2012) "On the trusted use of large-scale personal data", *Bulletin of the IEEE Technical Committee on Data Engineering*, vol. 35 (4), pp. 5-8.
[70] de Montjoye, Y., Shmueli, E., Wang, S. and Pentland, A. (2014) "openPDS" protecting the privacy of metadata through SafeAnswers", *PLOS One*, https://doi.org/10.1371/journal.pone.0098790
[71] Ward, P. (2015) "Hub of All Things", *Digital Leaders*, pp. 58-59, British Computer Society.
[72] Poikola A., Kuikkaniemi, K. and Honko, H. (2015) *MyData - A Nordic Model for Human-Centered Personal Data Management and Processing*, http://urn.fi/URN:ISBN:978-952-243-455-5

Both MyData and HAT expose *raw* data to applications and thus fail to limit the potential 'function creep'[73] that currently characterises data processing in the IoT and results in personal data flowing unfettered around the ecosystem. Both openPDS and the IoT Databox put severe constraints on the flow of data, minimising it to the results of data processing. While this too has the potential to expose users in ways they might not wish, e.g., through running multiple applications from a developer that allows them to build rich profiles from an array of returned results, the risk can be mitigated, e.g., through applications that monitor app usage and notify users as to the potential inferences that can be drawn from combined processing results.

Although openPDS is aligned with the European Commission's reform of the data protection rules, the IoT Databox seeks to respond directly to the external accountability requirement mandated by GDPR. In doing so it provides users *and* developers with a more extensive set of tools for GDPR compliant data processing in the IoT. It does so by implementing A29 WP IoT recommendations, general and specific, in the design and use of a local controlling and processing entity that empowers end-users. In responding to the interpersonal accountability requirement, and enabling individuals to manage one another's access to data, the IoT Databox also moves beyond the dominant 'individual-centric' approach adopted by openPDS and other solutions. As Crabtree and Mortier point out, most personal data do not belong to a single individual but are *relational* and thus *social* in nature, especially in the IoT where connected devices are embedded in the fabric and furniture of buildings.[74] The ability to share devices, data, and applications within and between homes, and to collectively as well as individually manage data processing, is also a unique feature of the IoT Databox model. The model turns upon what we might call 'computational accountability', insofar as it surfaces opaque machine-to-machine interactions and the social actors on whose behalf they operate. Manifests in particular make a currently invisible and unobtrusive digital ecology 'visible and rational',[75] making specific socio-technical data processing arrangements, implicating specific connected devices, data controller's and their processors, accountable to individuals and available to local control. In doing so, the IoT Databox reflexively enables legal accountability to be built into the Internet of Things.

**DEMONSTRABLY DOING ACCOUNTABILITY IN THE INTERNET OF THINGS**
The IoT Databox is able to respond to the 2-part compliance and demonstration requirements of the GDPR Accountability Principle. It enables substantive compliance through a variety of measures, documented below. It also communicates that compliance with users through various forms of account provided by the *manifest* and the *dashboar*d. In this section, we explore how the IoT Databox meets the accountability requirements detailed in Table 1.

---

[73] A29 WP223 (2014) Opinion 8/2014 on Recent Developments on the Internet of Things, http://ec.europa.eu/justice/data-protection/article-29/documentation/opinion-recommendation/files/2014/wp223_en.pdf

[74] Crabtree, A. and Mortier, R. (2015) "Human data interaction: historical lessons from social studies and CSCW", *Proceedings of ECSCW*, pp. 1-20, Oslo, Springer.

[75] Dourish, P. and Button, G. (1998) "On technomethodology: foundational relationships between ethnomethodology and system design", *Human Computer Interaction*, vol. 13 (4), pp. 395-432.

**Requirement 1. Limiting Initial Data Collection**

Limiting data collection in the IoT is challenging insofar devices are intentionally designed to collect extensive information in order to provide contextually aware services (such as fine-grained heating management or home security). Limiting initial data collection also sits uneasily with commercial models underpinning IoT technologies, which seek to repurpose data.[76] Nonetheless, only collecting what is functionally necessary for an IoT device, application or service to operate, i.e., for specific purposes, is clearly mandated by GDPR. The IoT Databox limits data collection through a number of architectural design choices. 1) Raw data from IoT devices is stored in separate containers on-the-box. 2) Computing is taken to the data at the edge of the network, rather than data to centralised computing, through the design of apps. 3) Apps cannot be installed on the box without a manifest being completed by the data subject, which includes the selection of data collection settings. Thus, the IoT Databox enables compliance by implementing an alternative architecture for the IoT. The demonstration of compliance occurs through data minimisation – the architecture constrains data distribution to the results of queries – and granular choice mechanisms embedded in the manifest.

**Requirement 2. Limitations on International Data Transfer**

Just where in the world data is distributed to requires attention under GDPR, particularly if they are transferred outside Europe or an adequate third country. The architecture of the IoT Databox negates many (but by no means all) questions to do with international data transfer, in locating data on-the-box, i.e., on a physical device situated in the data subject's home. The question of where the data is stored potentially arises if the data subject runs a virtual back up of the data held on the physical box. However, in this case, and unlike current cloud-based PDMSs, the decision of where to put the data is ultimately up to the user. The IoT Databox thus enables data subjects to control where their data is stored. The question of where the data is stored also arises with respect to the results of queries run on-the-box. The outcomes of analytics *can* travel internationally. The manifest again plays an important role here in providing an account to the user not only of what will be done with their data but by whom, including (in the condensed layer) other recipients of the data, and, where applicable, relevant adequacy decisions or safeguards. The manifest also makes it accountable (in the short layer) whether or not data will be taken off the box, and if online access is provided to the data subject if so. That some degree of risk is occasioned by apps that take data off the box is also clearly flagged to the data subject by an app's risk rating. While the IoT Databox cannot prevent data being taken off the box, it can incentivise a reduction in data transfer through the risk rating mechanism, with apps that take data off the box, pass it on to further recipients, particularly those located outside the EU, and/or which do not provide online access being severely rated. Further demonstrations of compliance may be achieved in the future through the use of machine-readable add-ons to data transfers such as blockchain, smart contracts or sticky policies.[77]

**Requirement 3. Responding to the Spectrum of Control Rights**

In enforcing local control, the IoT Databox negates the spectrum of control rights to some extent, for insofar as data stays on-the-box there is no need for access, rectification, objection, restriction, portability or erasure. External support for the spectrum of rights will still be required in cases where the results of local processing leave the box, though the raw data is retained on-the-box and the specific processing operations performed on the raw data are logged for audit, or where IoT devices first export data to the cloud. Insofar as APIs make data available to the data subject or data is provided in a common machine-readable format, as per the right to data portability, then data subjects may store such data on the box and reuse it for other purposes. We note, however, a particular shortcoming of the right to portability, namely that it does not cover statistical inferences, which are common to IoT data processing. The spectrum of control rights does not necessarily prevent potential *harms* that stem from lack of control over inferences, as opposed to raw data, then.[78] The IoT Databox seeks to address this situation by incentivising local processing in the app construction and app store processes, which emphasize and make potential risk accountable to *developers* as well as data subjects. Thus, the SDK prompts reflection on the potential risks created by an app in the very course of its construction, and the app store makes these risks (if they are not addressed) publically visible, as does the app manifest. Apps that do not take data off-the-box are also clearly accredited on the app store. Local control over processing is further enabled through the ability to preview the results of processing prior to distribution, should data be taken off the box. The ultimate sanction enables data subjects to immediately terminate data processing and revoke access.

---

[76] See, for example, "Roomba maker may share maps of users' homes with Google, Amazon or Apple", https://www.theguardian.com/technology/2017/jul/25/roomba-maker-could-share-maps-users-homes-google-amazon-apple-irobot-robot-vacuum

[77] Konstantinos Christidis and Michael Devetsikiotis, "Blockchains and Smart Contracts for the Internet of Things," *IEEE Access* 4 (2016): 2292–2303, doi:10.1109/ACCESS.2016.2566339; Siani Pearson and Marco Casassa-Mont, "Sticky Policies: An Approach for Managing Privacy across Multiple Parties," *Computer* 44, no. 9 (September 2011): 60–68, doi:10.1109/MC.2011.225.

**Requirement 4. Guaranteeing Greater Transparency Rights**
GDPR establishes a mandate for opening up the opaque IoT and providing more transparent information to data subjects. Transparency is key to enabling control rights insofar as data subjects cannot action them without knowing who controllers are, and what they are doing to their data. The IoT Databox takes significant steps to increase transparency in surfacing machine to machine interactions and the social actors on whose behalf they operate. The manifest clearly plays an important role in this respect, with the multi-layered notice approach scaffolding information depending on intended audience, providing legal and technical information as well as 'user-friendly' accounts of apps and their data processing operations. The app store also enhances transparency in providing social feedback through a commonly understood 'rating' mechanism. Further to this, the dashboard provides the data subject with a view of data processing in real time, including notifications of what the apps are doing and a preview function that enables data subjects to see the results of data processing before distribution. The ability to inspect how data flows through an app and how some action or decision is arrived at also speaks to the transparency and accountability in algorithms requirement, as detailed in Article 22 GDPR. Nonetheless, and despite these measures to enable accountability, the provision of meaningful information in 'plain' language about processing, as required by Article 12 GDPR, remains a challenge insofar as it cannot be enforced technologically.

**Requirement 5. Ensuring Lawfulness of Processing**
It is a condition of GDPR that the lawful grounds upon which processing stand be made accountable to data subjects. While a number of legitimate grounds are possible in law, the IoT Databox provides *only* for consent as the legal basis for processing and operating on the platform. The use of multi-layered notices, coupled with the underlying architecture, plays a key role here in turning the various permissions implicated in consenting to data processing (particularly the data sources used, and the frequency of sampling and reporting) into enforceable data processing policies managed by the IoT Databox arbiter. As noted above, the data subject may withdraw consent at any time and terminate processing. The limitation of this approach is that IoT devices which first export data to the cloud and subsequently make it available to data subjects via APIs render data subjects reliant on third party terms and conditions, including other grounds for processing.

---
[78] Urquhart et al Data Portability

**Requirement 6: Protecting Data Storage and Security**
Given the poor standard of security that currently effects the IoT, there is a lot of work to be done in safeguarding data and satisfying quick data breach notifications. GDPR clearly pushes towards technical approaches to doing this.[79] A combination of IoT Databox approaches assist with this requirement. First, the IoT Databox distributes data across containers. Containers are also manifold, with each unique data source having not only a container of its own but a potential array of containers associated with it produced as the result of data processing (e.g., an app may process data from temperature, humidity, and air quality sensors and create a new data store that holds data on environmental conditions in the home). This containerised approach exploits a 'honeycomb' rather than 'honey pot' approach (unlike centralised cloud servers) forcing an attacker to 'hack' each container to access data, which is itself encrypted at rest. This approach does not prevent attacks on online data sources (i.e., IoT devices which first export data to the cloud), but it does make attacking data on-the-box extremely challenging. Furthermore, the arbiter component (in line with consent permissions) regulates and manages which apps get to run on-the-box, what data are accessed, and what processing operations are allowed to run on the data. The IoT Databox also logs all data processing operations, including data export, for audit. These logs may be analysed to identify potential data breaches, though challenges exist in working out how to best achieve this (e.g., can breaches be detected automatically or is manual identification required by an expert).

**Requirement 7: Articulating and Responding to Processing Responsibilities**
GDPR encourages the adoption of mechanisms for data controllers to articulate their responsibilities, which are key to increasing public trust in the IoT. The SDK provides a relatively lightweight way of articulating compliance challenges to developers in the construction of apps and accompanying manifests. IoT developers, especially those in SMEs and start-ups, may lack the organisational resources to understand and respond to these challenges,[80] a situation more broadly confounded by the knowledge and skills deficit confronting organisational actors (e.g., lawyers lacking technical knowledge and technologists lacking legal knowledge). The SDK plays a key role in helping developers understand key accountability requirements and the potential impact of apps on data subjects, articulating a range of potential risks, legal, technical and social, occasioned by app development and incentivises developers to address them before uploading apps onto the app store. The app store in turn conveys the risks occasioned by an app to the public, an app's un/verified status, and accredits apps that do not take data off-the-box, thus certifying that the *highest* degree of responsibility is exercised in data processing by app developers.

| Accountability Requirement | IoT Databox Compliance Measure |
|---|---|
| **1. Limiting Initial Data Collection** | - The IoT Databox architecture situates data processing at the edge of the network in the data subject's environment, enables the data subject to control external access to data via app |

---

[79] Article 32 GDPR
[80] Lachlan Urquhart, "White Noise from the White Goods? Conceptual & Empirical Perspectives on Ambient Domestic Computing," in *Future Law*, ed. Edina Edwards, Lilian; Schafer, Burkhard; Harbinja (Edinburgh: Edinburgh University Press, 2017), 39, doi:10.2139/ssrn.2865738.

| | |
|---|---|
| | manifests that provide granular choice encoded as enforceable data processing policies on-the-box, and constrains data distribution to the results of processing. |
| **2. Limitations on International Data Transfer** | - Insofar as data is not taken off-the-box by an app, the requirement is negated.<br>- Insofar as data is taken off-the-box by the user backing up data in the cloud, the user retains control over where the data is placed for storage.<br>- Insofar as data is taken off-the-box by app, the manifest provides the data subject with the information required by Article 13 GDPR in the condensed layer.<br>- The app manifest also flags data transfer as a risk in the short layer, which increases in direct relation to access, further recipients and their location, adequacy decisions and safeguards. |
| **3. Responding to the Spectrum of Control Rights** | - Insofar as data is not taken off-the-box by an app, the requirement is negated.<br>- Insofar as data is taken off-the-box by an app, the spectrum of rights is outside the control of the IoT Databox, though the raw data is retained on-the-box and the specific processing operations performed on the raw data are logged for audit.<br>- Insofar as data is taken off-the-box by an app and online access is provided, any data made available to the data subject (either via APIs or in common machine-readable formats) may be imported into the IoT Databox for other processing as per the right to data portability.<br>- As the right to data portability does not include statistical inferences derived from data processing, the IoT Databox seeks to incentivise the construction of apps that do not take data off-the-box through risk rating mechanisms in the SDK, the app store, and the app manifest, and through accreditation.<br>- The IoT Databox allows data subjects to preview the results of data processing prior to distribution, to terminate data processing at any time, and to revoke access. |
| **4. Guaranteeing Greater Transparency Rights** | - The app manifest surfaces machine to machine interactions and the social actors on whose behalf they operate and processing takes place.<br>- Manifests exploit a multi-layered approach to make data processing accountable in legal, technical and 'user-friendly' terms.<br>- The dashboard enables the data subject to view data processing in real time, including notifications of processing operations, previews of the results of processing, and inspection of how results were arrived at as per Article 22.<br>- The app store exploits a commonly understood social 'rating' mechanism to enhance transparency. |
| **5. Ensuring Lawfulness of Processing** | - Data processing on the IoT Databox operates on the grounds of consent only.<br>- Consent is provided for by the app manifest, with specific permissions to access data sources and granular choices over data sampling and reporting frequencies being encoded as enforceable processing policies by the IoT Databox arbiter.<br>- Consent may be withdrawn and processing terminated by the data subject at any time. |
| **6. Protecting Data Storage and Security** | - The IoT Databox stores data in a distributed array of containers, which encrypt data at rest.<br>- The arbiter component regulates and manages app access to data stores and data processing operations, based permissions set by the data subject.<br>- All data processing operations, including data export, are logged for audit. |
| **7. Articulating and Responding to Processing Responsibilities** | - The SDK articulates compliance challenges to IoT developers in the course of app and manifest construction.<br>- The app store articulates risks associated with an app to the public, including legal, technical and social risks and an app's verified status.<br>- The app store also accredits apps that do not take data off-the-box, certifying the highest level of responsibility in data processing. |

**Table 2.** How the IoT Databox meets Accountability Requirements

The IoT Databox thus provides a prima facie example of what a demonstrably compliant account might 'look like' from a technological perspective, and to whom accountability is demonstrated: primarily IoT developers/data controllers and data subjects individually and en masse, but also to a potential range of other parties from technical experts to regulators and supervisory authorities, whose inquiries might be facilitated by the transparent and auditable characteristics of the IoT Databox platform. The account is articulated through a distinctive set of socio-technical *interactional arrangements* provided by the platform, which as can be seen in Table 2 demonstrably respond to the accountability requirements of GDPR. It is thus in

*doing* being accountable in the course of constructing, publishing and using IoT Databox apps that the *demonstration* of compliance is achieved. That demonstration reflexively tackles the various regulatory problems of the IoT, which occasion the need for accountability in the first instance. Thus, the IoT Databox makes an opaque technological infrastructure visible, providing clear consent mechanisms rendering data processing legible and enabling data subjects to control the flow of data, and also provides oversight on machine to machine communications along with what the data is being used for, where, and by whom. In making data processing in the IoT accountable, the IoT Databox seeks to go beyond remedying existing problems with the IoT however. Ultimately it seeks to foster a culture of accountability that results in widespread data minimisation, and entirely localised processing wherever possible, engendering widespread trust in the IoT.

> *"Data protection must move from 'theory to practice' … accountability based mechanisms have been suggested as a way [to] … implement practical tools for effective data protection."*[81]

**ACKNOWLEDGEMENT**
This work was supported by the Engineering and Physical Sciences Research Council [grant numbers EP/M001636/1, EP/N028260/1, EP/M02315X/1]

---

[81] WP173 (2010) "Opinion 3/2010 on the principle of accountability", *Article 29 Data Protection Working Party*, http://ec.europa.eu/justice/data-protection/article-29/documentation/opinion-recommendation/files/2010/wp173_en.pdf